\documentclass[english,letterpaper]{aa}
\usepackage{epsf,amsfonts,amssymb,graphicx,fancyheadings,caption}
\usepackage{natbib}
\usepackage{babel} 
\bibpunct{(}{)}{;}{a}{}{,}

\begin{document}

\title{EROS 2 photometry of probable R Coronae Borealis stars in the Small Magellanic Cloud
\thanks{Based on observations made with the CNRS/INSU MARLY telescope at the European
Southern Observatory, La Silla, Chile.
}}

\author{
P. Tisserand \inst{1},
J.B.~Marquette \inst{2},
J.P.~Beaulieu \inst{2},
P. de Laverny \inst{3},
\'{E}.~Lesquoy \inst{1,2},
A.~Milsztajn \inst{1},
C.~Afonso \inst{1,4},
J.N.~Albert \inst{5},
J.~Andersen \inst{6},
R.~Ansari \inst{5},
\'{E}.~Aubourg \inst{1},
P.~Bareyre \inst{1,4},
F.~Bauer \inst{1},
G.~Blanc\inst{1,7,8},
X.~Charlot \inst{1},
C.~Coutures \inst{1},
F.~Derue \inst{5,9},
R.~Ferlet \inst{2},
P.~Fouqu\'{e} \inst{10,11},
J.F.~Glicenstein \inst{1},
B.~Goldman \inst{1,4},
A.~Gould \inst{4,12},
D.~Graff \inst{13},
M.~Gros \inst{1},
J.~Haissinski \inst{5},
C.~Hamadache \inst{1},
J.~de Kat \inst{1},
T.~Lasserre \inst{1},
L.~Le Guillou \inst{1},
C.~Loup \inst{2},
C.~Magneville \inst{1},
B.~Mansoux \inst{5},
\'{E}.~Maurice \inst{14},
A.~Maury \inst{11},
M.~Moniez \inst{5},
N.~Palanque-Delabrouille \inst{1},
O.~Perdereau \inst{5},
L.~Pr\'{e}vot \inst{14},
Y.~Rahal \inst{5},
N.~Regnault \inst{5},
J.~Rich \inst{1},
M.~Spiro \inst{1},
A.~Vidal-Madjar \inst{2},
L.~Vigroux \inst{1},
and S.~Zylberajch\inst{1}
}

\institute{CEA, DSM, DAPNIA, Centre d'\'{E}tudes de Saclay, 91191 Gif-sur-Yvette
Cedex, France \and
Institut d'Astrophysique de Paris, INSU CNRS, 98bis boulevard Arago,
75014 Paris, France \and
Cassiop\'ee UMR 6202, Observatoire de la C\^ote d'Azur, BP4229, 06304 
Nice cedex 4, France \and
Coll\`{e}ge de France, Physique Corpusculaire et Cosmologie, IN2P3
CNRS, 11 place Marcelin Berthelot, 75231 Paris Cedex 05, France \and
Laboratoire de l'Acc\'{e}l\'{e}rateur Lin\'{e}aire, IN2P3 CNRS,
Universit\'{e} de Paris-Sud, 91405 Orsay Cedex, France \and
Astronomical Observatory, Copenhagen University, Juliane Maries Vej
30, 2100 Copenhagen, Denmark \and
Osservatorio Astronomico di Padova, INAF, vicolo dell'Osservatorio
5, 35122 Padova, Italy \and
Universit\'{e} Paris 7 Denis Diderot, 2 place Jussieu, 75005 Paris,
France \and
LPNHE, IN2P3 CNRS and Universit\'{e}s Paris 6 \& Paris 7, 4 place
Jussieu, 75252 Paris Cedex 05, France \and
Observatoire Midi-Pyr\'en\'ees, UMR 5572, 14 avenue Edouard Belin, 31400 Toulouse, France \and
European Southern Observatory, Casilla 19001, Santiago 19, Chile \and
Department of Astronomy, Ohio State University, Columbus, OH 43210,
U.S.A. \and
Department of Math and Science, U.S. Merchant Marine Academy, Kings Point, NY 11024, U.S.A. \and
Observatoire de Marseille, 2 place Le Verrier, 13248 Marseille Cedex
04, France }

\offprints{J.B. Marquette; \email{marquett@iap.fr}}

\date{Received ; Accepted}

\abstract{EROS 2 (Exp\'erience de Recherche d'Objets Sombres) conducted a
  survey of the SMC between July 1996 and February 2003 in two EROS
  broad-band colours, $V_\mathrm{E}$ and $R_\mathrm{E}$. The photometric data of 4.2 million stars have
  been searched for behaviour typical of R Coronae Borealis (RCB) candidates
  such as drastic changes in magnitude. Five objects have been found, four of
  them being catalogued in the Simbad database as \object{RAW 21}, \object{RAW 233},
\object{RAW 476}, and \object{[MH95] 431} with confirmed
carbon-rich atmospheres, characteristic of RCB. 
From the EROS~2 light curve  of RAW 21 and its spectrum reported
by \citet{2003MNRAS.344..325M}, we confirm that it is the
first RCB to be found in the SMC. The other objects are new RCB candidates
with absolute luminosity and colour close to those found for RCBs
in the LMC. We propose that 2 of them are DY~Per-like RCBs.

}

\authorrunning{P. Tisserand et al. (EROS coll.)}
\titlerunning{R Cor Bor candidates in the SMC}

\maketitle

\section{Introduction}

As described by \citet{1996PASP..108..225C} in a detailed review, R Coronae
 Borealis (RCB) stars are carbon-rich supergiants in a phase of rapid
 evolution.
They are born-again giants for which two evolutionary scenarios have been
 proposed: either a merger of two white dwarfs or a final He-shell flash in a
 post-AGB star. They exhibit extreme and irregular changes in brightness, up
 to 8 magnitudes in visible light, caused by the obscuration of the stellar
surface by newly formed dust clouds.
They are rare objects: about 50 are known in the Galaxy although population 
estimates are of a few thousand.
Furthermore, no Galactic RCBs have accurate distances
so that the absolute luminosities of these variables are poorly known.
This is why it is important to detect RCB objects in external galaxies, such
as the Magellanic Clouds (MC), whose distances are known. 
Studying RCBs in such low-metallicity systems is also crucial in order
to better understand their evolutionary status and the time spent
in the RCB phase.
Thanks to the work
of the MACHO collaboration on the LMC
\citep{1996ApJ...470..583A, 2001ApJ...554..298A}, the number
of detected RCB variables in this galaxy has increased from 3 to 17;
they estimate that the total number of LMC RCBs is less than 100.
However, \citet{2001ApJ...554..298A} stated in the same study that none
were found in the search performed in their SMC fields.

More recently, \citet{2003MNRAS.344..325M} reported a spectroscopic study of a
sample of $\sim$ 2300 MC carbon stars from which it appeared that six of them
show C$_{2}$ bands but weak or absent CN bands, a typical spectral
signature of RCBs.
They claimed that one of them is the first RCB candidate detected in the SMC, namely RAW 21. 
This acronym RAW comes from the catalog of carbon stars by 
\citet{1993A&AS...97..603R} who spectrophotometrically identified 1707 
objects in the SMC.
\\ The present paper reports on the mining of the EROS 2 (Exp\'erience de Recherche d'Objets Sombres) database in a search for RCB candidates in the SMC.
Results of this study are available at the URL {\tt http://eros.in2p3.fr/Variables/RCB/RCB-SMC.html}.

\section{Observational data \label{observations}}

The EROS 2 project made use of a 1-meter telescope located at La Silla Observatory, Chile, to extensively search for the baryonic dark matter of the Galactic halo. The method was based on the microlensing effect as described by \citet{2003A&A...404..145A}, \citet{2001A&A...373..126D}, \citet{2000A&A...355L..39L}, and \citet{1998A&A...332....1P}. The observations were performed between July 1996 and February 2003 with two wide field cameras (0.7\degr ~in right ascension $\times$ 1.4\degr ~in declination) behind a dichroic cube splitting the light beam into two broad passbands. The so-called ``blue'' channel (420--720 nm, hereafter magnitudes $V_\mathrm{E}$)
overlapped the $V$ and $R$ standard bands while the ``red'' one (620--920 nm, hereafter magnitudes $R_\mathrm{E}$) roughly matched the $I$ standard band. Each camera was a mosaic of eight 2048 $\times$ 2048 CCDs with a pixel size of 0.6\arcsec ~on the sky. \\
For the observations, the SMC was separated into ten fields, each of them having a surface of $\sim$ 1 deg$^2$. These fields were called sm001 to sm010 and labelling of the stars was identical to that done by \citet{2002A&A...389..149D}.
The photometry of individual images and the reconstruction of the light curves were processed using the Peida package which has been specifically developped for the EROS experiment \citep{1996VA.....40..519A}.
An estimation of the photometric accuracy is discussed by \citet{2002A&A...389..149D}.
Recently all available SMC data were used to recompute the entire sample of light curves. Over 5.6 million sources were detected on our template images.  These images are constructed by co-adding the 15 best seeing images of each field obtained between January 1997 and May 2001.  Once artifacts and the faintest objects were eliminated, we analysed more than 4.2 million light curves.  Stars were imaged on 650 epochs for outer fields and 950 epochs for inner fields. \\

\section{Mining the EROS database \label{mining.sect}}

As stated by \citet{1996PASP..108..225C} RCB light curves in the optical exhibit the following properties : initial decline of 3 to 6
magnitudes in about 50--100 days (at irregular intervals); fast or slow recoveries and/or multiple declines follow; final recovery may
 be slow. Guided by these properties, we defined criteria to search for a few RCB candidates among millions of objects. First we
extracted the most variable stars by correlating light curves taken through the ``red'' channel and the ``blue'' channel: objects
with a correlation coefficient greater than 0.5 were kept (about 5 \% of the stars). After this step a medium flux was calculated in
each colour as $F_\mathrm{med} = 1/2 \times (F_\mathrm{max} + F_\mathrm{min})$ where $F_\mathrm{max}$ and $F_\mathrm{min}$ are
the maximum and minimum
fluxes, respectively. The quantities $F_\mathrm{max}$ and $F_\mathrm{min}$ were determined as the averages of 10 points at maximum
and at minimum, respectively, chosen after elimination of the 5 most extreme points in both maximum and minimum domains. The number
of times a light curve crosses $F_\mathrm{med}$ by at least 3 consecutive points in either ``red'' or ``blue'' colour was used to
select stars for which this quantity was lower than 12. Remaining objects were then kept if
$2.5 \times \mathrm{log}_{10}(F_\mathrm{max}/F_\mathrm{min}) >$ 2.0 mags in either ``red'' or ``blue'' colour. This yielded
a sample
of 429 stars which were visually inspected to eliminate Miras or nova-like light curves and photometric artifacts, while still
retaining stars for which a rapid decrease (or increase) of luminosity is observed.  

\section{Detection efficiency}
There was until now no confirmed RCB star in the SMC. We have thus estimated our RCB detection efficiency by applying the cuts
described in the previous section to the 17 known LMC RCB stars, as listed in Table 1 of \citet{2001ApJ...554..298A}. One of
these 17 stars (HV 12842) is outside our fields and is not considered further. Using the light curves measured by
\citet{2001ApJ...554..298A} between mid-1992 and the end of 1999, we find that 14 stars out of 16 satisfy
our selection criteria. Among the 2 rejected stars, one
is a DY Per. We intend to report on these LMC stars in a future publication. \\
In addition, we have searched our LMC database for these 16 stars light curves and we have checked whether they satisfy
the selection criteria, based on EROS 2 data alone (July 1996 to February 2003). Seven of them are selected, none of
them being a DY Per. None of the remaining 9 stars from \citet{2001ApJ...554..298A} showed large enough variations
satisfying our selection criteria during the EROS 2 observing period.
This allows a rough estimate of the probability to detect an RCB candidate with about 450 photometric measurements over 6.5 years
at about 50 \%. (This number should be typical for the LMC ; in the SMC, where we have 650 to 950 epochs, our detection
efficiency might be higher.)

\section{The EROS2 SMC RCB candidates}
After the visual inspection of the 429 selected light curves, five objects remain which are listed in Table \ref{RCB.tab}. Four of these candidates were retrieved in the Simbad database. The abbreviation [MH95] in Table \ref{RCB.tab} refers to the spectroscopic survey of 1185 carbon stars in the SMC by \citet{1995A&AS..113..539M}. The J2000 coordinates of the objects are those measured on the EROS 2 template images with an average accuracy  of 1\arcsec.

\begin{table*}
\caption{Coordinates of the EROS 2 SMC RCB candidates.
\label{RCB.tab}}
\medskip
\centering
\begin{tabular}{llcll}
\hline
\hline
Simbad identifier & EROS2-RCB identifier & J2000 coordinates \\
\hline
RAW 21		& J003747-733902(sm0102l20592b) & 00:37:47.07  -73:39:02.1 \\
RAW 233		& J004407-724416(sm0077k11497b)$^a$ & 00:44:07.45  -72:44:16.3 \\
RAW 476		& J004822-734104(sm0014k11612b) & 00:48:22.87  -73:41:04.7 \\
{[MH95]} 431	& J004014-741121(sm0106m19412b) & 00:40:14.65  -74:11:21.2 \\
n/a		& \object{J005718-724235(sm0067m28134b)}$^b$ & 00:57:18.12  -72:42:35.3 \\
\hline
\multicolumn{5}{l}{$^a$ MACHO identifier: 208.15571.60}\\
\multicolumn{5}{l}{$^b$ MACHO identifier: 207.16426.1662; hereafter identified as sm0067m28134b}\\
\end{tabular}
\end{table*}

\begin{table*}
\caption{2MASS and unpublished DENIS infrared data for the EROS 2 SMC RCB candidates.
\label{RCB.IRdata}}
\medskip
\centering
\begin{tabular}{llllllll}
\hline
\hline
Identifier	& JD Epoch & $J_{\mathrm{2MASS}}$ & $H_{\mathrm{2MASS}}$ & $K_{\mathrm{2MASS}}$ & $I_{\mathrm{DENIS}}$ & $J_{\mathrm{DENIS}}$ & $K_{\mathrm{DENIS}}$ \\
\hline
RAW 21		& 2\,451\,033.9	& 17.06	& 15.23	& 13.49	&	&	&       \\
		& 2\,450\,412.6	& 	&	&	& 15.72	& 14.65 & 12.43 \\
		& 2\,451\,394.9	& 	&	&	& 14.89	& 13.66 & 11.87 \\
RAW 233		& 2\,451\,034.7	& 13.31 & 12.03 & 11.30 &	&	&       \\
		& 2\,450\,040.6	& 	&	&	& 15.29	& 13.43 & 11.30 \\
		& 2\,450\,418.6	& 	&	&	& 14.91	& 13.13 & 11.05 \\
		& 2\,451\,048.8	& 	&	&	& 15.40	& 13.37 & 11.27 \\
RAW 476		& 2\,451\,034.7	& 13.12 & 12.69 & 12.11 &	&	&       \\
		& 2\,450\,418.6	& 	&	&	& 13.70	& 13.11 & 11.98 \\
		& 2\,451\,035.9	& 	&	&	& 13.76	& 13.11 & 12.02 \\
		& 2\,451\,050.8	& 	&	&	& 13.99	& 13.32 & 12.15 \\
		& 2\,451\,416.8	& 	&	&	& nd	& nd 	& 13.16 \\
{[MH95]} 431	& 2\,451\,033.9	& 14.63 & 13.54 & 12.95 &	&	&       \\
		& 2\,450\,413.6	& 	&	&	& 14.73	& 13.33 & 12.06 \\
		& 2\,450\,414.6	& 	&	&	& 14.82	& 13.30 & 11.78 \\
sm0067m28134b	& 2\,451\,034.8	& 13.88 & 12.81 & 11.73	&	&	&       \\
		& 2\,450\,432.6	& 	&	&	& nd	& nd	& 13.71 \\
\hline
\multicolumn{8}{l}{nd = not detected}\\
\end{tabular}
\end{table*}

\begin{figure}
	\resizebox{\hsize}{!}{\includegraphics{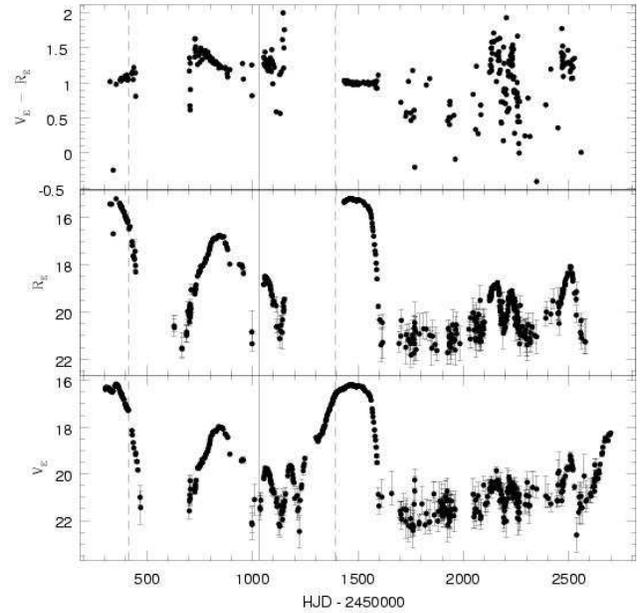}}
	\caption{EROS light curves of RAW 21. The vertical solid line
	  indicates the epoch of the 2MASS observation while dashed lines are DENIS observations.}
	\label{RAW21.fig}
\end{figure}

\begin{figure}
	\resizebox{\hsize}{!}{\includegraphics{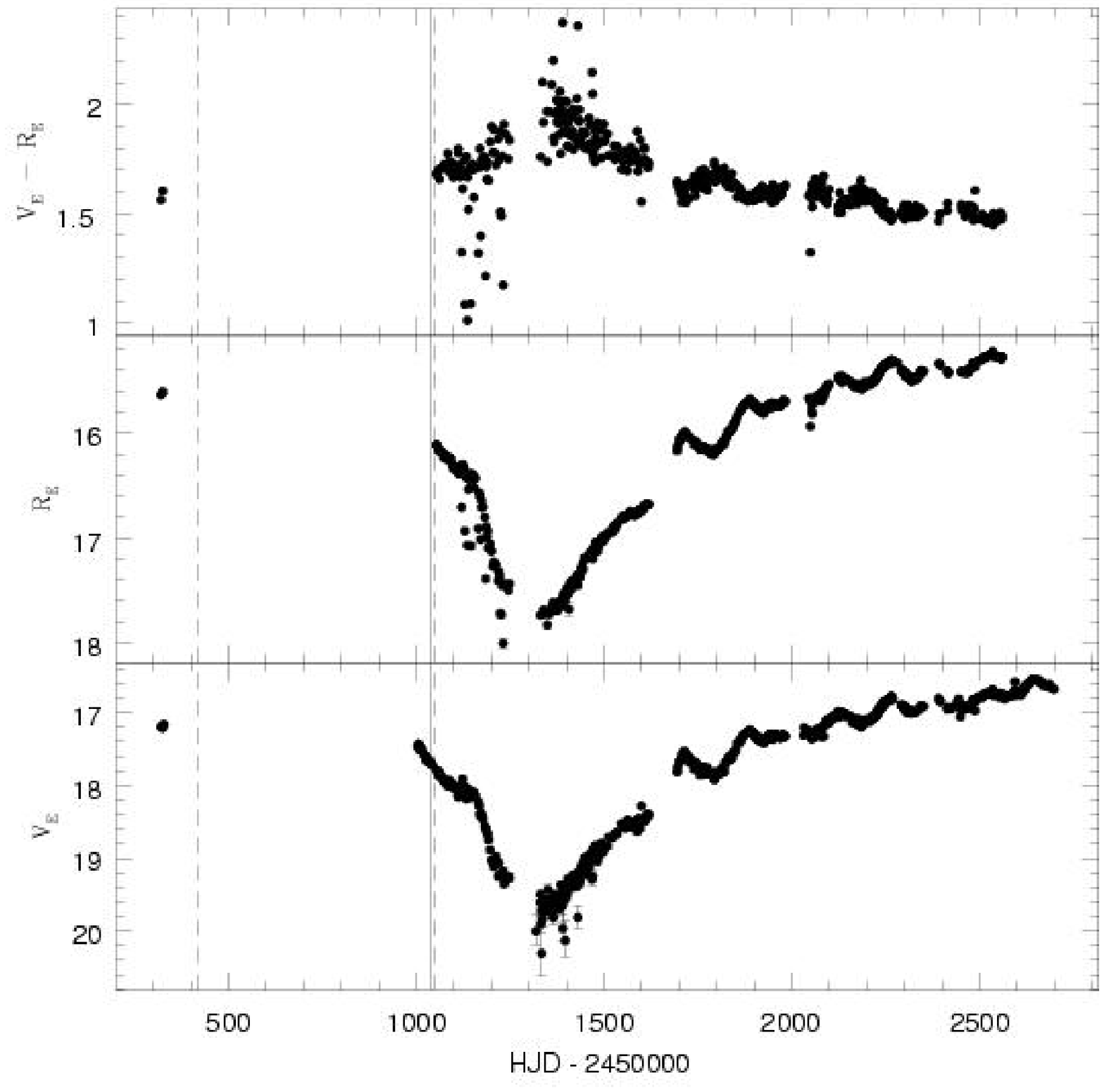}}
	\caption{EROS light curves of RAW 233. The vertical solid line indicates the epoch of the 2MASS observation
while dashed lines are DENIS observations.}
	\label{RAW233.fig}
\end{figure}

\begin{figure}
	\resizebox{\hsize}{!}{\includegraphics{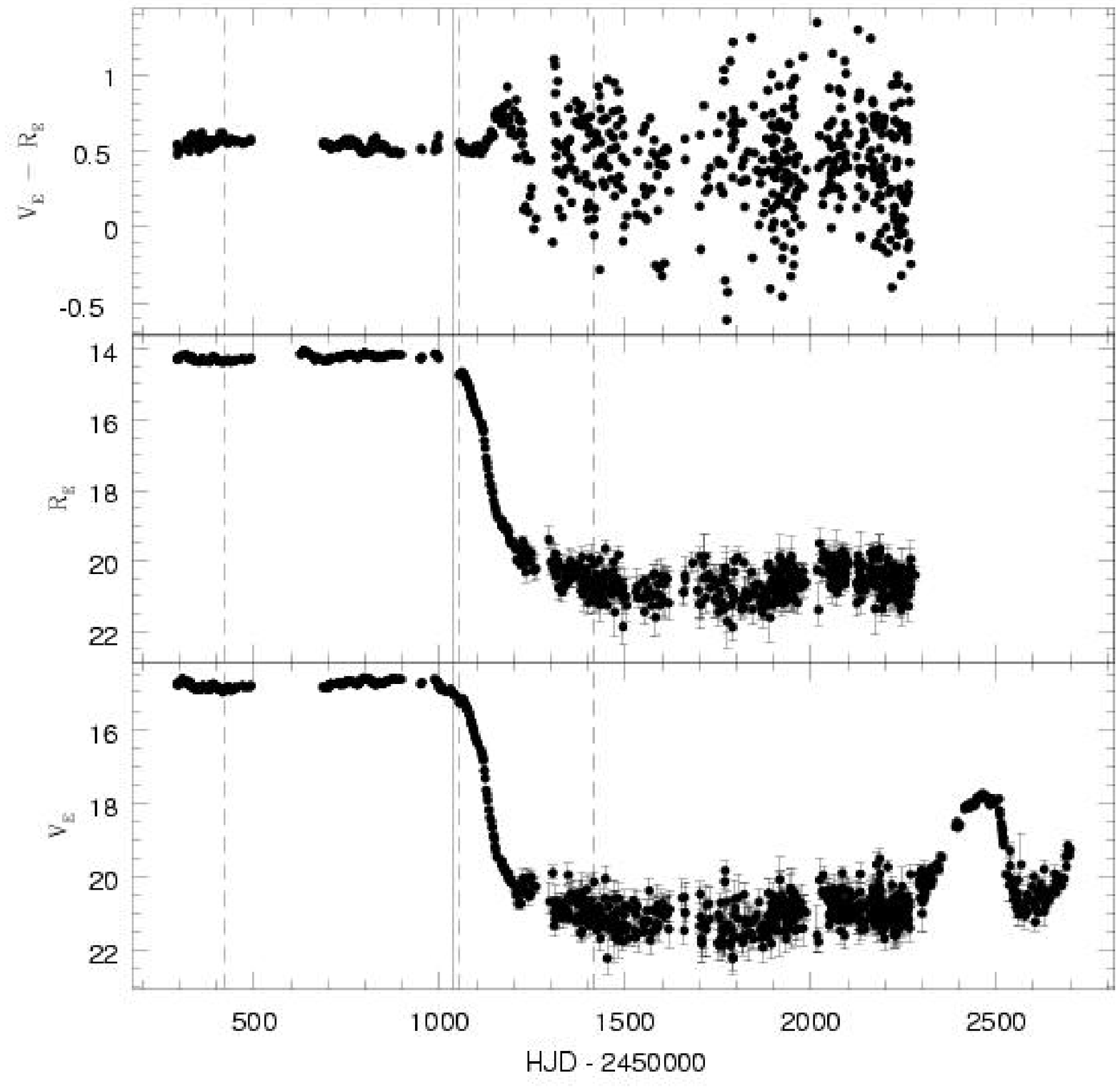}}
	\caption{EROS light curves of RAW 476. The vertical solid line indicates the epoch of the 2MASS observation
while dashed lines are DENIS observations.}
	\label{RAW476.fig}
\end{figure}

\begin{figure}
	\resizebox{\hsize}{!}{\includegraphics{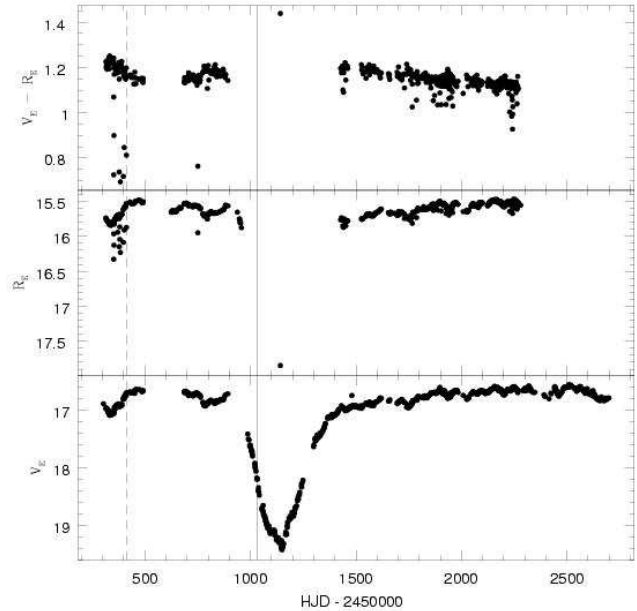}}
	\caption{EROS light curves of [MH95] 431. The vertical solid line indicates the epoch of the 2MASS observation
while the dashed line is the DENIS observation.}
	\label{MH95_431.fig}
\end{figure}

\begin{figure}
	\resizebox{\hsize}{!}{\includegraphics{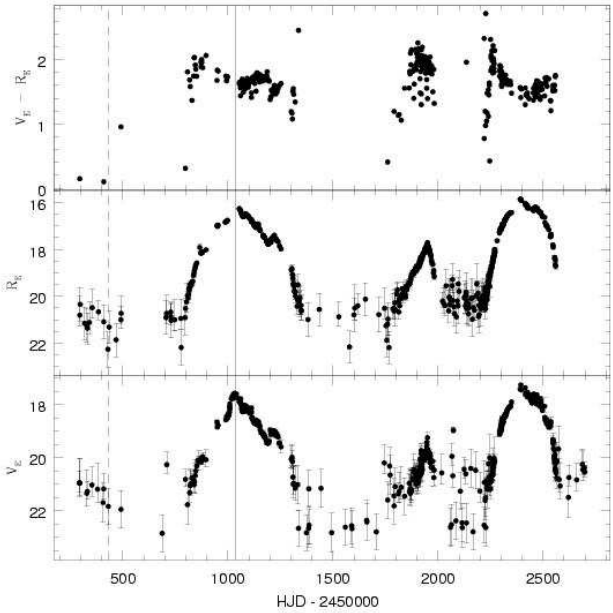}}
	\caption{EROS light curves of sm0067m28134b. The vertical solid line indicates the epoch of the 2MASS observation
while the dashed line is the DENIS observation.}
	\label{sm0067m28134b.fig}
\end{figure}

Figures \ref{RAW21.fig} to \ref{sm0067m28134b.fig} display the light curves of
the RCB candidates in both $V_\mathrm{E}$ and $R_\mathrm{E}$ colours together
with the temporal evolution of the colour index $V_\mathrm{E} -
R_\mathrm{E}$.
Also indicated by vertical lines are the epochs of 2MASS and (mostly unpublished) DENIS observations.
Data for RAW 233 and sm0067m28134b are available from the
Lightcurve Retrieval on the MACHO webserver\footnote{{\tt
    http://store.anu.edu.au:3001//cgi-bin/lc.pl}} for the coordinates given in
Table \ref{RCB.tab}. Some CCDs of the EROS ``red''
camera had technical problems from time to time. This is the reason for the
wide gaps in the $R_\mathrm{E}$ data, which
unfortunately coincide for some objects with interesting magnitude changes
detected with the ``blue'' camera. However it remains clear that for all
objects the light and colour
variations are large and erratic. For the sake of clarity, error bars have been omitted
on the colour index in Figs. \ref{RAW21.fig} to \ref{sm0067m28134b.fig},
and points for which the error on either $V_\mathrm{E}$ or $R_\mathrm{E}$ is greater
than 1 mag have been removed.

\begin{figure}
	\resizebox{\hsize}{!}{\includegraphics{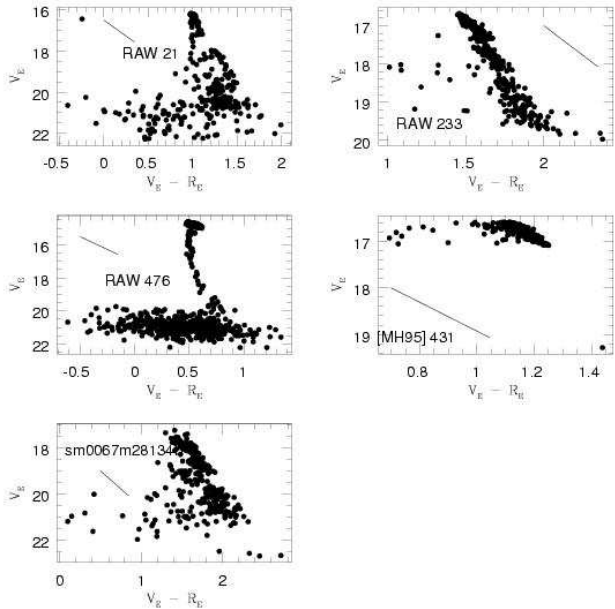}}
	\caption{Colour--magnitude diagram of the RCB candidates. The solid lines give the direction of the reddening vectors whose determination is discussed in the text.}
	\label{CMD.fig}
\end{figure}

\begin{figure}
	\resizebox{\hsize}{!}{\includegraphics{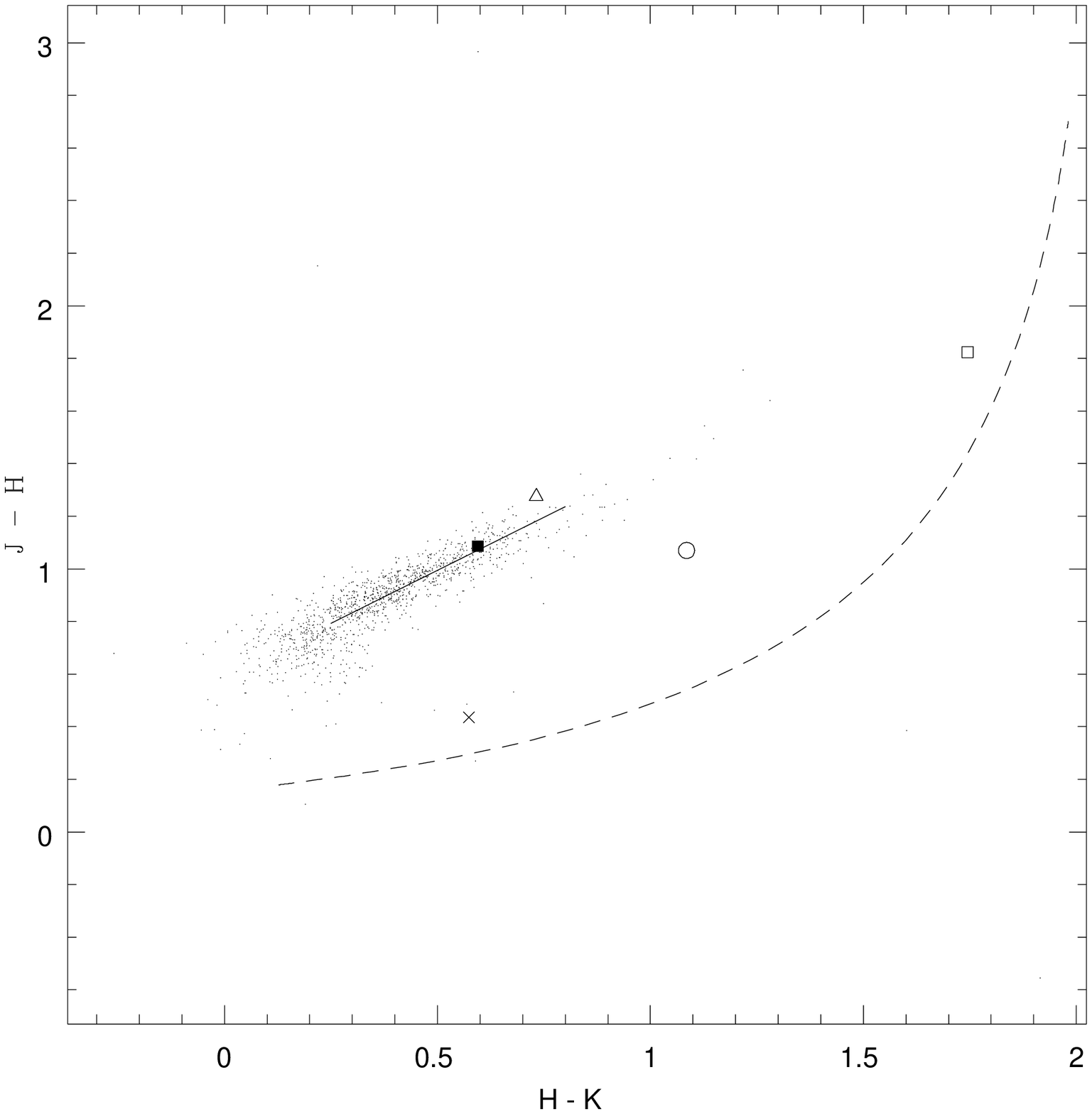}}
	\caption{$J - H$ versus $H - K$ two--colour diagram. Solid dots~:
	  2MASS data collected through VizieR for the 1185 carbon stars of the
	  catalog by \citet{1995A&AS..113..539M}; open square~: RAW 21; open
	  triangle~: RAW 233; cross~: RAW 476; filled square~: [MH95] 431;
	  open circle~: sm0067m28134b. The solid line represents the colours
	  of SMC carbon stars as parameterised by \citet{1991A&AS...91..425W}. The dashed line
	  is the locus of a combination of 5500~K (the ``star'') and 1000~K (the ``shell'')
	  blackbodies in various proportions ranging from all ``star'' (lower end)
	  to all ``shell'' (upper end), as calculated by \citet{1997MNRAS.285..339F}.}
	\label{colour.fig}
\end{figure}

Figure \ref{CMD.fig} shows the colour--magnitude excursion of each object. 
The solid lines indicate the direction of a reddening vector
corresponding to a standard extinction, which is probably a rough approximation
for a carbon-rich atmosphere. For the EROS passbands it is determined according to Eqs. (1) and (3)
of \citet{1989ApJ...345..245C} with effective wavelengths equal to 0.626
$\mu$m and 0.788 $\mu$m for $V_\mathrm{E}$ and $R_\mathrm{E}$,
respectively. The quantities $A_\mathrm{V}$ and $E(B-V)$ for SMC are taken
from the NASA/IPAC Extragalactic Database (NED). 

Figure \ref{colour.fig} shows the positions of the RCB candidates in a near infrared two--colour
diagram constructed with the 2MASS data obtained through VizieR for the carbon
stars of the catalog by \citet{1995A&AS..113..539M}. The solid line
corresponds to a model by \citet{1991A&AS...91..425W} giving the colour of the
SMC carbon stars. It should be noticed that the colours of 
RAW 476, sm0067m28134b and RAW 21 seem to be linearly correlated as are the
LMC RCBs of \citet{2001ApJ...554..298A}. The locations of these stars are also compared
to the locus of a combination of two blackbodies, e.g. a 5500 K star and a 1000 K dust shell
in various proportions ranging from all star (lower end of the dashed line)
to all shell (upper end of the dashed line), as proposed by \citet{1997MNRAS.285..339F}.

\subsection{RAW 21}
From its carbon-rich spectra with strong C$_2$ bands but very weak CN
bands \citep{2003MNRAS.344..325M}, this star was proposed as the first 
RCB candidate ever detected in the SMC. 
But its variability data were too sparse 
until now to definitively conclude on its nature. 
Thanks to the EROS~2 light curve revealing huge brightness variations, we 
can claim
that RAW~21 is indeed the first RCB discovered in the SMC.

Figure \ref{CMD.fig} shows that a standard extinction does not explain what is 
observed during the phases of luminosity changes.
In addition, a drastic change in $J$
and $K$ magnitudes is observed when comparing the values from 2MASS and DENIS
given in Table \ref{RCB.IRdata}. 
This reflects the different epochs of the 2MASS and DENIS observations.
The optical EROS~2 data collected at these epochs also
reveal that RAW~21 was brighter when observed by DENIS. Furthermore,
\citet{2003MNRAS.344..325M} assumed that this RCB
was much fainter in June~2000 than in January~2000 since they failed to
collect a spectrum in June. EROS~2 data also confirm such a variation.

Finally, the
temporal evolution of $J - K$ suggests that this object has experienced a major 
episode of local extinction which is also observed in the optical domain 
as shown on Fig.~\ref{RAW21.fig}. This is consistent with the classical
interpretation for the RCB brightness declines assuming that they
are caused by successive obscurations by dust clouds.

\subsection{RAW 233}
The EROS~2 lightcurve shown in Fig.~\ref{RAW233.fig} is a very
good example of RCB brightness variations with (i) a large decline
of about 2--3~mags, (ii) a larger decline at shorter wavelengths
compatible with dust obscuration, and (iii) regular or semi-regular pulsations
around maximum light with amplitudes of a few tenths of magnitude and 
pseudo-periods of a few tens of days.
The MACHO light curves collected between JD $\sim$ 2\,448\,800 and JD $\sim$ 2\,451\,400
exhibit an earlier episode of slow magnitude decrease and increase with 
superimposed similar low amplitude variations as those observed on
Fig. \ref{RAW233.fig}. 

The colour diagram on Fig. \ref{CMD.fig} for this star shows a decrease which
is steeper than the extinction vector. 
The almost symmetrical brightness decrease and increase and the position 
of this object on Fig. \ref{colour.fig} are comparable with that
of DY Per stars (cooler and fainter Galactic RCB with slower
symmetrical declines) 
indicated by \citet{2001ApJ...554..298A} and close to the 
line giving the colour of the SMC carbon stars by
\citet{1991A&AS...91..425W}. 

From our photometric data and its already known carbon-rich nature, 
we therefore propose that RAW~233 is the first RCB of the DY~Per-type detected
in the SMC. Its cooler effective temperature must
be confirmed spectroscopically.

\subsection{RAW 476}
The evolution of this carbon-rich object on Fig.~\ref{RAW476.fig} is 
the most spectacular with a decline of almost 7 magnitudes after a 
long episode of relative stability. We also point out that very 
low-amplitude pulsations are 
detected at maximum light and the decline is deeper at shorter wavelengths.
The extinction in $V_\mathrm{E}$ on Fig. \ref{CMD.fig} is practically
 colour--independent. The Combined General Catalog of Variable Stars
 \citep{1998GCVS4.C......0K} reports for this object a magnitude at 
maximum brightness of 15.4 and an irregular type of variability.
Its RCB nature might thus be definitively established if
this star has recovered to this maximum brightness after the EROS~2
observations and if its spectrum exhibits strong C$_2$
and weak CN bands.

\subsection{[MH95] 431}
The carbon star [MH95]~431 is another very good RCB candidate. 
It exhibits
on Fig. \ref{MH95_431.fig} low amplitude and colour-dependent magnitude 
changes at maximum light and a huge decline in $V_\mathrm{E}$.
However, the lack of data in $R_\mathrm{E}$ during the main decline rules out 
any definitive conclusion on a reddening at this epoch. 
Notice however that the points at JD $\sim$ 2\,451\,130 are real
and are compatible with the dust obscuration scenario. 
Here again low amplitude variations are observed and the two-colour diagram of
Fig. \ref{colour.fig} shows an almost
perfect agreement with the colour of SMC carbon stars of
\citet{1991A&AS...91..425W}. Again, this object is located in the same region 
of the diagram as the DY~Per stars of \citet{2001ApJ...554..298A}.
Its almost symmetrical decline is indeed compatible
with a possible DY~Per nature; its cooler temperature should
be checked spectroscopically.

\subsection{sm0067m28134b}
\begin{figure}
	\resizebox{\hsize}{!}{\includegraphics{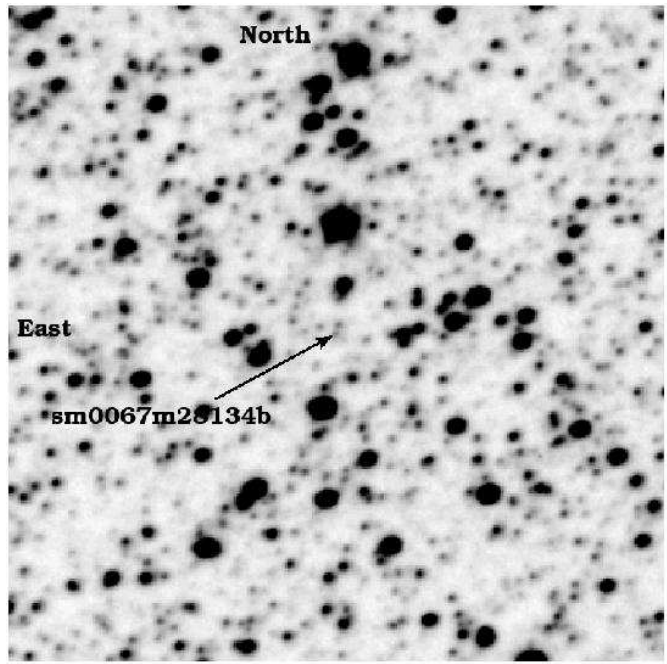}}
	\caption{Finding chart of sm0067m28134b which is indicated by the arrow. The size of the frame
is 168\arcsec $\times$ 168\arcsec. North is up, East is left.}
	\label{Champ.fig}
\end{figure}

This star is the only one which is not included in a published catalog of
carbon stars. This could be explained by the fact that it is very faint
on the EROS template image as can be seen on the finding chart of
Fig. \ref{Champ.fig}. On the other hand there is at this location a quite
bright 2MASS source (see Table \ref{RCB.IRdata}). In the same way the GSC 2.2
catalog available through VizieR yields a source with a F photographic
magnitude of 17.04 while visible magnitudes are not determined. This strongly
suggests the presence of a star embedded in a screening dust shell and visible
at irregular intervals. The MACHO light curve is available in one colour only; it
displays an earlier and even larger variation
of about 6 mags between the epochs JD $\sim$ 2\,449\,300 and
JD $\sim$ 2\,450\,000.
If its RCB nature is confirmed spectroscopically, this object
has undergone a large number of obscurations by dust clouds
revealing an interesting phase of dust ejections.

\begin{figure}
	\resizebox{\hsize}{!}{\includegraphics{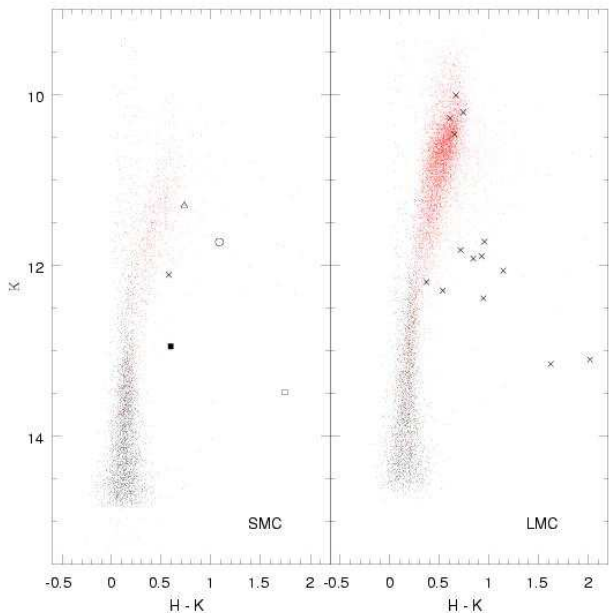}}
	\caption{$K$ versus $H - K$ colour--magnitude diagram. Left panel:
	SMC; the symbols for our 5 RCB candidates are the same as in Fig. \ref{colour.fig}; black solid
	dots are discussed in the text; red solid dots are the carbon stars of
	the catalog by \citet{1995A&AS..113..539M}. Right panel: LMC; crosses
	represent RCB stars given by \citet{2001ApJ...554..298A} in their
	Table 1 for which 2MASS data exist; black solid dots are discussed in the text; red solid dots
	are 7609 carbon stars extracted from the catalog of
	\citet{2001A&A...369..932K}. }
	\label{diagHK.fig}
\end{figure}

\subsection{Comparison with LMC objects}
The steepest luminosity decrease observed for our 5 candidates
are of order 0.01 mag/day for the 2 possible DY Per, 
and 0.06 - 0.12 mag/day for the other three; 
these values are very similar to those given by \citet{2001ApJ...554..298A}.

Figure \ref{diagHK.fig} shows the 2MASS $K$ versus $H - K$
colour--magnitude diagram of our RCB candidates (left panel)
compared to the RCB stars found in the LMC (right panel). For 3 LMC objects out of 17, no obvious 2MASS
counterpart was found and they are omitted in the figure. The solid black dots are stars chosen
as follows : for each RCB candidate a query was made on the 2MASS catalog
through VizieR to extract 1000 sources within a radius of 10\arcmin~around the
object and with a JHK photometric quality flag equal to AAA. Finally, the red dots
represent either the catalog of \citet{1995A&AS..113..539M} (for SMC) or that
of \citet{2001A&A...369..932K} (for LMC). 

From this figure, it appears that both the SMC \& LMC 
carbon star populations show similar patterns and that our SMC
RCB candidates are located in similar
regions with respect to these carbon stars.
Notice that the four LMC DY Per stars proposed by \citet{2001ApJ...554..298A}
are well separated from the other objects (four upper crosses in the right
panel of Fig. \ref{diagHK.fig}) while no such difference is seen on the SMC
part of the figure. This indicates that the DY Per-like position of RAW 233
and [MH95] 431 on Fig. \ref{colour.fig} is not confirmed on
Fig. \ref{diagHK.fig}. Clearly spectroscopic measurements are necessary to
disentangle this question.

Finally, we have estimated the absolute visual magnitude at maximum light
($M_\mathrm{V}$ in the Johnson system) of the RCB candidates
in the SMC. 
They are presented in Table~\ref{Mv.tab}
together with their colour at maximum light. 
For that purpose, we assumed a distance modulus of 18.9 \citep{2003MNRAS.339..157H}
and $A_\mathrm{V}$ = 0.123 (NED), 
constant over the whole SMC field covered by EROS~2. We neglected the circumstellar extinction 
around each RCB. 
The standard $V$ and I magnitudes are estimated within a 10 \% accuracy
for each field and CCD by matching the EROS catalogue of stars with
those from DENIS and OGLE\footnote{Available at the URL\\
{\tt http://sirius.astrouw.edu.pl/$\sim$ogle/ogle2/smc\_maps.html}} \citep{1998AcA....48..147U}. \\
It should be noted that, for the first four stars in Table~\ref{Mv.tab}, we are confident
that the quoted magnitudes and colours correspond to the maximum luminosity. For
sm0067m28134b in contrast, we have evidence from the MACHO data that the maximum luminosity in
Fig. \ref{sm0067m28134b.fig} is about 2 mags fainter than the actual maximum.

RAW~21 and RAW~476 have absolute luminosities
compatible with the faintest ($M_\mathrm{V} \sim -2.5$) 
and the brightest ($M_\mathrm{V} \sim -5$)
RCB of the LMC respectively.
It can also be seen that the faintest RCB candidates
are the reddest as already noticed for LMC RCB (see 
\citet{2001ApJ...554..298A}).
The two DY~Per candidates of the SMC (RAW~233 and  [MH95]~431)
are the coolest SMC candidates
and have an absolute luminosity slightly fainter than, but still compatible with,
that of LMC DY~Per stars ($M_\mathrm{V} \sim -2.5$).
Finally, conclusions on sm0067m28134b are more uncertain as EROS 2 did not observe this
star at maximum light, while MACHO likely observed it but only in one passband.

To summarize, we find that all the RCB candidates in the SMC
follow the classical relationship between absolute magnitude
and effective temperature for the RCB variables and
have absolute luminosities close to those found in the LMC.

\begin{table}
\caption{Estimated absolute luminosity and colour at maximum light
of the RCB candidates 
in the SMC. \label{Mv.tab}}
\medskip
\centering
\begin{tabular}{lcc}
\hline
\hline
RCB candidate & $M_\mathrm{V}$ &  ($V - I$) \\
\hline
RAW 21                 & -2.5 & 1.9 \\
RAW 233	               & -2.1 & 2.1 \\
RAW 476	               & -4.5 & 1.0 \\
{[MH95]} 431           & -2.1 & 2.0 \\
sm0067m28134b          & -1.3 & 2.4 \\
\hline
\end{tabular}
\end{table}

\section{Summary}
A search has been performed for RCB candidates in the SMC among 4.2
millions light curves of the EROS~2 database. 
After applying various filters described in Sect. \ref{mining.sect}, five
objects have been identified as possible stars undergoing the RCB phase of
evolution (large declines, infrared excess at minimum light
and semi-regular variations at maximum).
Four of them being previously known as carbon stars,
their RCB nature is reinforced.

One of them, RAW~21, being already known as exhibiting strong
C$_2$ bands but very weak CN-bands, is therefore the first confirmed
RCB found in the SMC with an absolute luminosity close to that
of the faintest known RCB in the LMC. On the other hand, RAW~476
is almost as bright as the brightest known RCB.

From their position in the $J - H$ versus $H - K$ two--colour diagram, 
their absolute luminosity
and their symmetrical decline in the lightcurve, the RCB
type of the stars 
RAW 233 and [MH95] 431 is possibly DY Per--like although 
no such evidence appears
on the $K$ versus $H - K$ colour--magnitude diagram.
 
With the data for RAW~21, RAW~476 and sm0067m28134b, a linear correlation is
 observed in the $J - H$ versus $H - K$ two--colour diagram. 
The SMC candidates have similar infrared colours with respect to carbon stars as do LMC stars.

Little can be said about the abundance of RCB stars with only five candidates.  Based on their study of galactic RCB stars, \citet{2001ApJ...554..298A} suggested that only $\sim 10^{-6}$ of all stars are RCB stars.  We find that in the SMC,
$0.7 - 1.7 \times 10^{-6}$ $(1 \sigma)$ of all stars display photometric features characteristic of RCB stars over the course
of 6.5 years. (Taking into account our estimated detection efficiency, the actual fraction of RCB stars may be up to twice higher.)
Thus, there is no significant detection of any environmental, historical, or metallicity dependence on the fraction of RCB stars.

Finally, we suggest that spectroscopic studies checking the relative strength
of the C$_2$ and CN bands be carried out to definitively
confirm the RCB nature of RAW~233, RAW~476, [MH95]~431 and sm0067m28134b.

\begin{acknowledgements}
We thank the anonymous referee for his/her useful comments.
This research has made use of VizieR and the Simbad database, operated at CDS, Strasbourg, France. This publication makes use of data products from the Two Micron All Sky Survey, which is a joint project of the University of Massachusetts and the Infrared Processing and Analysis Center/California Institute of Technology, funded by the National Aeronautics and Space Administration and the National Science Foundation.
The unpublished DENIS data have been kindly provided by Gary Mamon.
The DENIS project has been partly funded by the SCIENCE and the HCM plans
of the European Commission under grants CT920791 and CT940627. It is
supported by INSU, MEN and CNRS in France, by the State of
Baden-W\"urttemberg in Germany, by DGICYT in Spain, by CNR in Italy, by
FFwFBWF in Austria, by FAPESP in Brazil, by OTKA grants F-4239 and
F-013990 in Hungary, and by the ESO C\&EE grant A-04-046.
\end{acknowledgements}

\bibliographystyle{aa}
\bibliography{RCB}

\begin{thebibliography}{19}
\expandafter\ifx\csname natexlab\endcsname\relax\def\natexlab#1{#1}\fi

\bibitem[{{Afonso} {et~al.}(2003){Afonso}, {Albert}, {Alard}, {Andersen},
  {Ansari}, {Aubourg}, {Bareyre}, {Bauer}, {Beaulieu}, {Blanc}, {Bouquet},
  {Char}, {Charlot}, {Couchot}, {Coutures}, {Derue}, {Ferlet}, {Fouqu{\' e}},
  {Glicenstein}, {Goldman}, {Gould}, {Graff}, {Gros}, {Haissinski},
  {Hamadache}, {Hamilton}, {Hardin}, {de Kat}, {Kim}, {Lasserre}, {LeGuillou},
  {Lesquoy}, {Loup}, {Magneville}, {Mansoux}, {Marquette}, {Maurice}, {Maury},
  {Milsztajn}, {Moniez}, {Palanque-Delabrouille}, {Perdereau}, {Pr{\' e}vot},
  {Regnault}, {Rich}, {Spiro}, {Tisserand}, {Vidal-Madjar}, {Vigroux}, \&
  {Zylberajch}}]{2003A&A...404..145A}
{Afonso}, C., {Albert}, J.~N., {Alard}, C., {et~al.} 2003, \aap, 404, 145

\bibitem[{{Alcock} {et~al.}(2001){Alcock}, {Allsman}, {Alves}, {Axelrod},
  {Becker}, {Bennett}, {Clayton}, {Cook}, {Dalal}, {Drake}, {Freeman}, {Geha},
  {Gordon}, {Griest}, {Kilkenny}, {Lehner}, {Marshall}, {Minniti}, {Misselt},
  {Nelson}, {Peterson}, {Popowski}, {Pratt}, {Quinn}, {Stubbs}, {Sutherland},
  {Tomaney}, {Vandehei}, \& {Welch}}]{2001ApJ...554..298A}
{Alcock}, C., {Allsman}, R.~A., {Alves}, D.~R., {et~al.} 2001, \apj, 554, 298

\bibitem[{{Alcock} {et~al.}(1996){Alcock}, {Allsman}, {Alves}, {Axelrod},
  {Becker}, {Bennett}, {Clayton}, {Cook}, {Freeman}, {Griest}, {Guern},
  {Kilkenny}, {Lehner}, {Marshall}, {Minniti}, {Peterson}, {Pratt}, {Quinn},
  {Rodgers}, {Stubbs}, {Sutherland}, \& {Welch}}]{1996ApJ...470..583A}
{Alcock}, C., {Allsman}, R.~A., {Alves}, D.~R., {et~al.} 1996, \apj, 470, 583

\bibitem[{{Ansari}(1996)}]{1996VA.....40..519A}
{Ansari}, R. 1996, Vistas in Astronomy, 40, 519

\bibitem[{{Cardelli} {et~al.}(1989){Cardelli}, {Clayton}, \&
  {Mathis}}]{1989ApJ...345..245C}
{Cardelli}, J.~A., {Clayton}, G.~C., \& {Mathis}, J.~S. 1989, \apj, 345, 245

\bibitem[{{Clayton}(1996)}]{1996PASP..108..225C}
{Clayton}, G.~C. 1996, \pasp, 108, 225

\bibitem[{{Derue} {et~al.}(2001){Derue}, {Afonso}, {Alard}, {Albert},
  {Andersen}, {Ansari}, {Aubourg}, {Bareyre}, {Bauer}, {Beaulieu}, {Blanc},
  {Bouquet}, {Char}, {Charlot}, {Couchot}, {Coutures}, {Ferlet}, {Fouqu{\' e}},
  {Glicenstein}, {Goldman}, {Gould}, {Graff}, {Gros}, {Ha{\" i}ssinski},
  {Hamilton}, {Hardin}, {de Kat}, {Kim}, {Lasserre}, {Le Guillou}, {Lesquoy},
  {Loup}, {Magneville}, {Mansoux}, {Marquette}, {Maurice}, {Milsztajn},
  {Moniez}, {Palanque-Delabrouille}, {Perdereau}, {Pr{\' e}vot}, {Regnault},
  {Rich}, {Spiro}, {Vidal-Madjar}, {Vigroux}, \&
  {Zylberajch}}]{2001A&A...373..126D}
{Derue}, F., {Afonso}, C., {Alard}, C., {et~al.} 2001, \aap, 373, 126

\bibitem[{{Derue} {et~al.}(2002){Derue}, {Marquette}, {Lupone}, {Afonso},
  {Alard}, {Albert}, {Amadon}, {Andersen}, {Ansari}, {Aubourg}, {Bareyre},
  {Bauer}, {Beaulieu}, {Blanc}, {Bouquet}, {Char}, {Charlot}, {Couchot},
  {Coutures}, {Ferlet}, {Fouqu{\' e}}, {Glicenstein}, {Goldman}, {Gould},
  {Graff}, {Gros}, {Ha\&{\i}ssinski}, {Hamilton}, {Hardin}, {de Kat}, {Kim},
  {Lasserre}, {Le Guillou}, {Lesquoy}, {Loup}, {Magneville}, {Mansoux},
  {Maurice}, {Milsztajn}, {Moniez}, {Palanque-Delabrouille}, {Perdereau},
  {Pr{\' e}vot}, {Regnault}, {Rich}, {Spiro}, {Vidal-Madjar}, {Vigroux},
  {Zylberajch}, \& {The EROS collaboration}}]{2002A&A...389..149D}
{Derue}, F., {Marquette}, J.-B., {Lupone}, S., {et~al.} 2002, \aap, 389, 149

\bibitem[{{Feast}(1997)}]{1997MNRAS.285..339F}
{Feast}, M.~W. 1997, \mnras, 285, 339

\bibitem[{{Harries} {et~al.}(2003){Harries}, {Hilditch}, \&
  {Howarth}}]{2003MNRAS.339..157H}
{Harries}, T.~J., {Hilditch}, R.~W., \& {Howarth}, I.~D. 2003, \mnras, 339, 157

\bibitem[{{Kholopov} {et~al.}(1998){Kholopov}, {Samus}, {Frolov}, {Goranskij},
  {Gorynya}, {Karitskaya}, {Kazarovets}, {Kireeva}, {Kukarkina}, {Kurochkin},
  {Medvedeva}, {Pastukhova}, {Perova}, {Rastorguev}, \&
  {Shugarov}}]{1998GCVS4.C......0K}
{Kholopov}, P.~N., {Samus}, N.~N., {Frolov}, M.~S., {et~al.} 1998, in Combined
  General Catalogue of Variable Stars, 4.1 Ed (II/214A, available through
  VizieR).

\bibitem[{{Kontizas} {et~al.}(2001){Kontizas}, {Dapergolas}, {Morgan}, \&
  {Kontizas}}]{2001A&A...369..932K}
{Kontizas}, E., {Dapergolas}, A., {Morgan}, D.~H., \& {Kontizas}, M. 2001,
  \aap, 369, 932

\bibitem[{{Lasserre} {et~al.}(2000){Lasserre}, {Afonso}, {Albert}, {Andersen},
  {Ansari}, {Aubourg}, {Bareyre}, {Bauer}, {Beaulieu}, {Blanc}, {Bouquet},
  {Char}, {Charlot}, {Couchot}, {Coutures}, {Derue}, {Ferlet}, {Glicenstein},
  {Goldman}, {Gould}, {Graff}, {Gros}, {Ha{\i}ssinski}, {Hamilton}, {Hardin},
  {de Kat}, {Kim}, {Lesquoy}, {Loup}, {Magneville}, {Mansoux}, {Marquette},
  {Maurice}, {Milsztajn}, {Moniez}, {Palanque-Delabrouille}, {Perdereau},
  {Pr{\' e}vot}, {Regnault}, {Rich}, {Spiro}, {Vidal-Madjar}, {Vigroux},
  {Zylberajch}, \& {The EROS collaboration}}]{2000A&A...355L..39L}
{Lasserre}, T., {Afonso}, C., {Albert}, J.~N., {et~al.} 2000, \aap, 355, L39

\bibitem[{{Morgan} \& {Hatzidimitriou}(1995)}]{1995A&AS..113..539M}
{Morgan}, D.~H. \& {Hatzidimitriou}, D. 1995, \aaps, 113, 539

\bibitem[{{Morgan} {et~al.}(2003){Morgan}, {Hatzidimitriou}, {Cannon}, \&
  {Croke}}]{2003MNRAS.344..325M}
{Morgan}, D.~H., {Hatzidimitriou}, D., {Cannon}, R.~D., \& {Croke}, B.~F.~W.
  2003, \mnras, 344, 325

\bibitem[{{Palanque-Delabrouille} {et~al.}(1998){Palanque-Delabrouille},
  {Afonso}, {Albert}, {Andersen}, {Ansari}, {Aubourg}, {Bareyre}, {Bauer},
  {Beaulieu}, {Bouquet}, {Char}, {Charlot}, {Couchot}, {Coutures}, {Derue},
  {Ferlet}, {Glicenstein}, {Goldman}, {Gould}, {Graff}, {Gros}, {Haissinski},
  {Hamilton}, {Hardin}, {de Kat}, {Lesquoy}, {Loup}, {Magneville}, {Mansoux},
  {Marquette}, {Maurice}, {Milsztajn}, {Moniez}, {Perdereau}, {Prevot},
  {Renault}, {Rich}, {Spiro}, {Vidal-Madjar}, {Vigroux}, {Zylberajch}, \& {The
  EROS Collaboration}}]{1998A&A...332....1P}
{Palanque-Delabrouille}, N., {Afonso}, C., {Albert}, J.~N., {et~al.} 1998,
  \aap, 332, 1

\bibitem[{{Rebeirot} {et~al.}(1993){Rebeirot}, {Azzopardi}, \&
  {Westerlund}}]{1993A&AS...97..603R}
{Rebeirot}, E., {Azzopardi}, M., \& {Westerlund}, B.~E. 1993, \aaps, 97, 603

\bibitem[{{Udalski} {et~al.}(1998){Udalski}, {Szymanski}, {Kubiak},
  {Pietrzynski}, {Wozniak}, \& {Zebrun}}]{1998AcA....48..147U}
{Udalski}, A., {Szymanski}, M., {Kubiak}, M., {et~al.} 1998, Acta Astronomica,
  48, 147

\bibitem[{{Westerlund} {et~al.}(1991){Westerlund}, {Azzopardi}, {Rebeirot}, \&
  {Breysacher}}]{1991A&AS...91..425W}
{Westerlund}, B.~E., {Azzopardi}, M., {Rebeirot}, E., \& {Breysacher}, J. 1991,
  \aaps, 91, 425

\end{thebibliography}

\listofobjects

\end{document}